\documentclass[dvips]{acta}
\usepackage{supertabular,lscape, epsfig}
\usepackage{amssymb}
\usepackage{amsmath}
\usepackage{graphicx}
\usepackage{times}
\usepackage{verbatim}
\usepackage{epsfig}

\SetPages{0}{0}
\SetVol{00}{0000}

\begin{document}

\begin{Titlepage}
\Title{Carriers of 4964 and 6196 diffuse interstellar bands and environments dominated by  either CH or CH$^{+}$ molecules}
\Author{Weselak$^{1}$, T., Galazutdinov$^{2, 3}$, G.A., Sergeev$^{4}$, O., 
        Godunova$^{4}$, V., Ko{\l}os$^{5}$, R., Kre{\l}owski$^{6}$, J.}
{$^1$ Institute of Physics, Kazimierz Wielki University, Weyssenhoffa 11, 85-072 Bydgoszcz, Poland
e-mail: towes@gazeta.pl\\

$^2$Instituto de Astronomia, Universidad Catolica del Norte, Av. Angamos 0610, 
    Antofagasta, 1270709 Chile\\

$^3$ Pulkovo Observatory, Pulkovskoe Shosse 65,
       Saint-Petersburg 196140, Russia\\
 e-mail: runizag@gmail.com\\

$^4$  International Center for Astronomical, Medical and Ecological Research\\
Postal code and address: 27 Zabolotnoho Str., Kyiv 03680 Ukraine\\
e-email: sergeev@terskol.com; godunova@mao.kiev.ua\\

$^5$ Institute of Physical Chemistry, Polish Academy of Sciences\\
  Kasprzaka 44, 01-224 Warsaw, Poland\\
e-mail: robert.kolos@ichf.edu.pl\\

$^6$ Center for Astronomy, Nicolaus Copernicus University,
  Grudzi{\c{a}}dzka 5, Pl-87-100 Toru{\'n}, Poland \\
 e-mail: jacek@astri.uni.torun.pl  }   

\Received{Month Day, Year}
\end{Titlepage}

\Abstract{The analysis of radial velocities of interstellar
spectral features: CH, CH$^{+}$ as well as 4964 and 6196 diffuse
interstellar bands, seen in spectra of HD 151932 and 152233,
suggests that carrier of the former is spatially correlated with
CH while that of the latter -- with CH$^{+}$. A further analysis,
done in this paper and based on  the sample of 106 reddened OB
stars, partly confirms this suggestion, showing  that the CH
column density correlates indeed much better with the equivalent
width of the 4964 DIB than with that of the 6196 DIB. However, the
strengths of the 6196 DIB correlate only marginally better with
CH$^+$ than with CH. } 

{ISM: molecules}

\section*{Introduction}

Diffuse interstellar bands (DIBs) are numerous narrow to broad
absorption features mostly found in the visual  {and} near
infrared (4,000-10,000\AA) spectra of early type OB--stars.
Despite many observational efforts (Herbig 1995, Galazutdinov et
al. 2000, Weselak et al. 2000, Hobbs et al. 2008, and references
therein), DIBs remain one of the longest standing puzzles in
stellar spectroscopy. Since their discovery (Heger 1922) not a
single DIB carrier has been identified, although many possible
candidates were proposed (Herbig 1995). The coincidence of a weak
DIB at 5069~\AA\ with the gas-phase absorption of the linear,
centrosymmetric hydrocarbon chain cation HCCCCH$^+$ has been
recently reported by Kre{\l}owski et al. (2010). However, this
identification, as well as the recent assignment of two broad
diffuse bands near 4882 and 5450~\AA\ to the propadienylidene
(\textit{l}-C$_3$H$_2$) molecule (Maier et al., 2011a) have been
disputed (Kre{\l}owski et al. 2011; Maier et al. 2011b) and remain
uncertain. Recent results of Ehrenfreund et al. (2002), Cox and
Patat (2008) proved the existence of DIBs and molecular features
in spectra of LMC/SMC and M100 stars, respectively. However,
measurements of weak interstellar features are hindered by
relatively low signal--to--noise (S$/$N) inside their profiles.

The molecular origin of at least some DIBs seems likely, their
profiles being often complex and reminiscent of rotational
envelopes in molecular spectra (Sarre et al. 1995a, Kerr et al.
1998). The 6614 DIB contains three main components that have the
appearance of unresolved P, Q and R branches of an electronic
transition of a large molecule (see Herzberg 1950).  A
more or less similar pattern is observed also in the case of 5797
DIB, the strength of which correlates tightly with the CH column
density (Weselak et al. 2008b). This correlation supports also the
idea that carriers of some DIBs are molecules. Another important
piece of evidence is that the width of certain DIB can depend on
the temperature within an interstellar cloud -  as shown by
Ka{\'z}mierczak et al. (2009) who have found a correlation between
the FWHM of  the 6196 DIB and the rotational temperature of the
homonuclear C$_2$ molecule. An important conjecture here is that
some of the DIBs, namely those sensitive to  temperature of the
environment, may originate in centrosymmetric molecules, as the
lack of  electric dipole moment slows down the rotational
relaxation  (see Bernath 2005).

Methylidyne (CH),  first identified in the ISM by McKellar (1940),
is closely related to molecular hydrogen (H$_{2}$), as already
shown by Mattila (1986) and Weselak et al. (2004) and also with OH
(Weselak et al. 2009b, 2010b). The abundances of CH molecule are also well
correlated with those of NH (Weselak et al. 2009c).\\
On the other
hand, the column density of the corresponding cation, CH$^{+}$,
correlates very poorly with that of H$_{2}$ - indicating no
relation between the abundances of these two molecules (Weselak et
al. 2008a). The formation and existence of CH$^{+}$ in the ISM
remains  an unsolved problem (van Dishoeck and Black 1989, Gredel
et al. 1993, Sheffer et al. 2008 and references therein). One can
argue that environments dominated by the CH molecule (i.e. regions
where it is much more abundant than CH$^+$), and those where
CH$^+$ dominates, are well separated in space. This conclusion is
grounded in the publication of Allen (1994) which demonstrates
that radial velocities of CH and CH$^+$ may differ by as much as
7.3 km/s.

Spectral observations of highly-reddened early type stars offer an
excellent opportunity to probe large column densities of gas along
diverse lines of sight, to derive physical parameters of
corresponding environments and eventually unveil the nature of DIB
carriers. Identified and unidentified interstellar spectral
features should share the radial velocities if they  originate in
the same environments. In such a case, their intensities should be
correlated as well.

The aim of this work is to present the relations between the
column densities of simple diatomic molecules CH and CH$^{+}$ and
the equivalent widths of two narrow DIBs: 4964 and 6196~\AA. It is
certainly important to specify the physical parameters of
environments which host the DIB carriers. It may shed some light
on the conditions facilitating the formation$/$preservation of DIB
carriers. This is particularly interesting in the context of
interstellar CH and CH$^{+}$ synthesis paths,  presented by
Federman et al. (1982) and by Zsarg{\'o} and Federman (2003).

\section{The observational data}

The observations\footnote {Based on observations made with ESO
Telescopes at the La Silla or Paranal Observatories under programs
71.C-0367(A), 073.D-0609(A), 074.D-0300(A), 075.D-0369(A),
076.C-0431(B) and 082.C-0566(A).} were obtained during several
runs spanning the period 1999--2007, using BOES (b), FEROS (f),
UVES (U) and HARPS (H) \'echelle spectrographs. We also retrieved
fully reduced UVES (u) spectra of several objects listed in
Table~1 from the "Library of High-Resolution Spectra of Stars
across the Hertzsprung-Russell
Diagram" (Bagnulo et al. 2003,\\
http://www.sc.eso.org/santiago/uvespop). Spectra were reduced
using MIDAS and IRAF, as well as our own DECH code (Galazutdinov
1992), which provides all standard procedures for image and
spectra processing. The DECH code was employed during the final
data analysis.

The usage of several different software data reduction packages
allows us to get better control over proper dark subtraction,
flatfielding, or excision of cosmic ray spikes. We measured the
equivalent widths of  selected features using a Gaussian fit (see
Weselak et al. 2009a) in the spectra obtained using different
instruments. Some of our targets were observed several times which
allowed for the estimation of repeatability (it is not  likely
that two observers make the same error). In such cases the
measurements were averaged.
In the case of
spectral features that were unsaturated (EW $<$ 20 m\AA),
the following relation were applied (Weselak et al. 2009a)
to obtain the column density (in $cm^{-2}$):\

\begin{center}
$N = 1.13\times10^{20} \left( EW/ f \lambda^{2} \right)$,
\end{center}

\noindent where $EW$ is equivalent width of the line, $\lambda$
its wavelength (both in~\AA), and $f$ -- its oscillator
strength.\\ We calculated the column density of the CH$^{+}$
cation adopting, for an usaturated 4232~\AA\ band, the oscillator
strength \emph{f=}0.00545 reported by Larsson and Siegbahn (1983).
The same procedure has been applied in a former study by Weselak
et al. (2008a).  Whenever the 4232~\AA\ band was saturated,
measurements of the  CH$^{+}$ feature at 3957~\AA\ were performed,
using its recently  obtained \emph{f}-value of  0.00342 (Weselak
et al. 2009a). When estimating the column density of  CH$^{+}$, we
only used the bands of the total equivalent width lower than
20m\AA. In the case of two objects where Doppler--spliting is
evident (HD 152235 and 168607), we  also made use of the 3957~\AA\
band to estimate the column densities. In the case of HD 34078, we
relied on the unsaturated CH$^{+}$ (2,~0) band at 3745\AA,\ with
its recently published
\emph{f}-value of 0.00172 (Weselak et al. 2009a).\\
In the case of CH molecule we calculated column density based
on unsaturated CH A--X band at 4300~\AA\, adopting recent
 \emph{f}-value of  0.00506 (Weselak et al. 2014).
When the 4300~\AA\ feature was saturated we used both
CH B--X bands at 3886 and 3890~\AA\ (\emph{f}-values equal to
 0.00320 and 0.00213, respectively -- see Weselak et al. 2014).
The CH B--X bands at 3886 and 3890~\AA\  were used in the case of
HD's: 34078, 147889, 154368 and 204827 respectively.

\section{Results}

For this project we choose a sample of 106 reddened OB stars, for
which we collected the echelle spectra featuring CH and CH$^{+}$
bands, as well as the DIBs at 4964 and 6196 \AA. For each line of
sight. Tables 1 and 2 list a stellar HD/ BD number (observed by b
-- BOAO, f -- FEROS, H -- HARPS, U -- UVES), spectral type,
luminosity class, colour excess E(B-V) (in magnitudes), as
well as the column densities of   CH$^{+}$ and CH molecules (in
10$^{12}$ cm$^{-2}$ units, together with corresponding   error
estimates), and finally the equivalent widths (W$_{\lambda}$'s) of
diffuse\  bands, with their standard deviations.
 In Figure~1 we present, in the radial
velocity scale, the CH and CH$^{+}$ features, as  seen in a very
high S/N ratio (2,500) spectrum of HD 149757 ($\zeta$Oph). It is
evident that radial velocities of both species are identical. At
the bottom of this figure we also show the two narrow DIBs at
4963.85 and 6195.97~\AA\ in the spectrum of HD 23180, taken from
the survey of Galazutdinov et al. (2000). A weak feature at
6194.76~\AA\ can be seen in the vicinity of 6196 DIB. Also the
radial velocities of 4964 and 6196~DIBs do not differ, as far as
their asymmetric profiles allow for central wavelengths
measurements.

To determine the DIB radial velocities we have used the rest
wavelength velocity frame based on the KI  line at 7698.974~\AA\
(Galazutdinov et al. 2000) toward HD 23180 where radial velocities
of all identified interstellar features are the same. This method
allows us to measure radial velocities of diffuse bands and of
narrow interstellar features with a very low uncertainty, smaller
than 0.3 km/s (see Fig. 1). HD 149757 is a moderately reddened,
popularly observed object, with no Doppler splittings of
interstellar features. The above statement is supported by the
observations of CH B-X lines (centered near 3886 and 3890~\AA) and
of the feature near 3957~\AA, belonging to the  CH$^{+}$ cation,
which proved useful for column density derivation when saturation
affected stronger bands (Weselak et al. 2008a, 2008b). We observed
$\zeta$Oph using UVES and HARPS spectrographs, and the result is
the same in both cases: no Doppler shift of any feature, relative
to any other one, is observed toward this target.

It is well known that the CH$^{+}$ cation needs different
interstellar conditions than CH to be formed (Federman 1982). In
many cases CH$^{+}$ features are Doppler-splitted, while those of
CH are not (see Pan et al. 2004). However, there are also lines of
sight where we observed the splitting in both CH and CH$^{+}$
bands. In Figure~2 we present a fragment of  HD~152233 spectrum,
measured with HARPS, with   CH and CH$^{+}$ (4300 and 4232~\AA,
respectively)\ features plotted in the radial velocity scale. The
CH$^+$ line is apparently Doppler splitted, and contains two
components; conversely,  only one component is detected for the
neutral CH molecule. At the bottom of Fig. 2 we  present the
spectral region of 4964 and 6196 DIBs, in the same velocity scale.
Intrinsic DIB wavelengths were those of Galazutdinov et al.
(2008). In the case of broader DIBs (see Fig. 1), the error of
radial velocity measurements is not larger than 1-2 km/s. It is
evident that the 6196 DIB shares its radial velocity with the
blue-shifted (-7.6$\pm$0.4 km/s) component of the CH$^{+}$  line,
while the 4964 DIB shares the radial velocity with CH. A similar
effect can be observed in the UVES spectrum of HD~151932, with the
blue--shift of CH$^+$ equal to  -6.1$\pm$0.2 km/s (Fig. 3).

\begin{figure}[ht!]
\epsfig{file=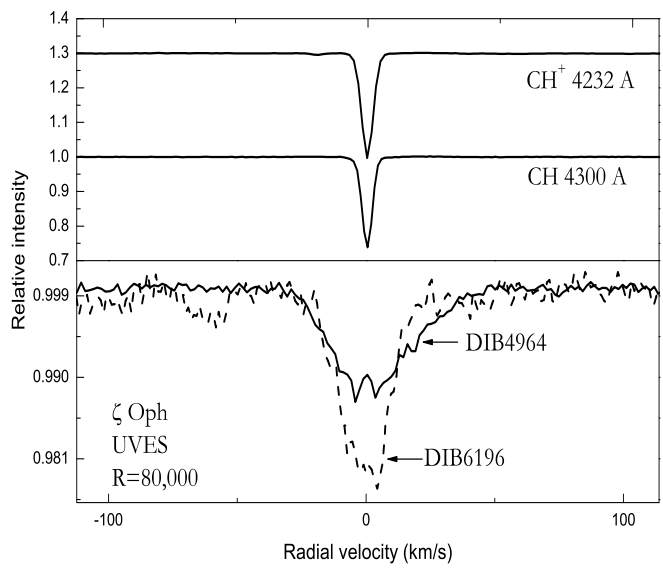, width=100mm, height=90mm, clip=} \caption{
Top panel: CH A-X  and CH$^{+}$ A-X lines centered near 4300 and
4232~\AA, respectively, as seen in the spectrum of HD 149757.
Common radial velocity scale. Bottom panel presents the DIBs at
4963.85 and 6195.97~\AA\ (Galazutdinov et al. 2000), in the same
radial velocity scale. A weak DIB at 6194.76~\AA\ can be
discerned. Difference in radial velocities is not higher than 0.3 km/s.
}
\end{figure}

\begin{figure}[ht!]
\epsfig{file=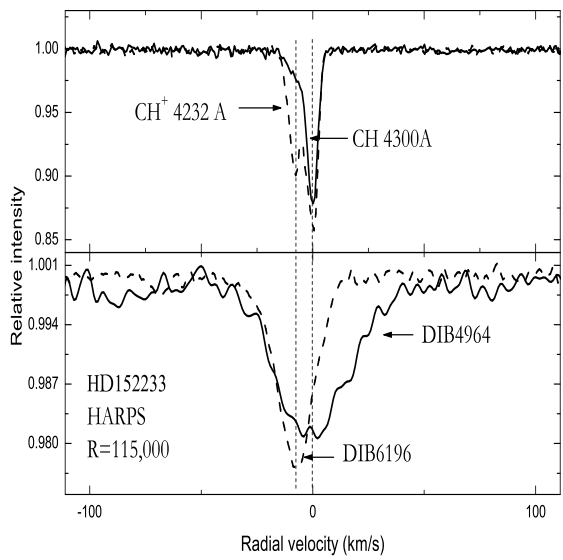, width=85mm, height=90mm, clip=} \caption{
The 6196 DIB is shown sharing the velocity of the second component
of CH$^+$, while 4964 DIB remains at the position of CH. The depth
of the 6196 DIB is scaled down by a factor of 2. It is
evident that the 6196 DIB shares its radial velocity with the
blue-shifted (-7.6$\pm$0.4 km/s), weaker component of the CH$^{+}$
line. }
\end{figure}

\vspace{-1.5cm}

\begin{figure}[ht!]
\epsfig{file=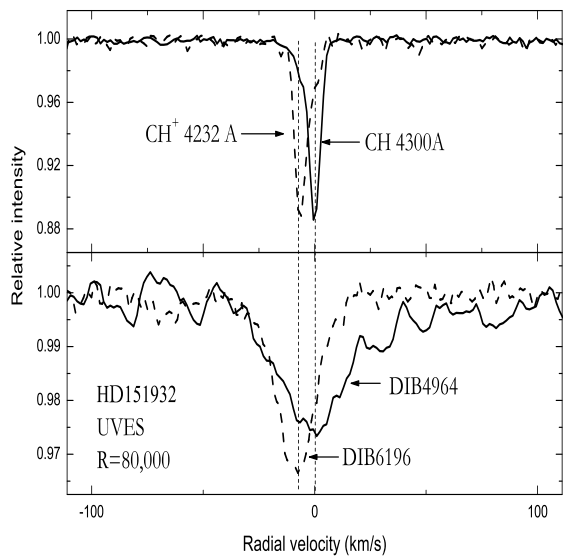, width=85mm, height=90mm, clip=} \caption{
Same as  Fig. 2 but for a different target observed with another
instrument. The depths of  4300~\AA\ CH line and of  6196 DIB are
scaled down by a factor of 2.
The 4964 DIB shares the radial velocity with CH
in the UVES spectrum of HD~151932, with the
blue--shift of CH$^+$ equal to  -6.1$\pm$0.2 km/s. }
\end{figure}

\clearpage

Such telltale Doppler shift analyses can at present,
unfortunately, not be accomplished for  a statistically meaningful
sample of cases, given the fact that velocity differences between
CH and CH$^{+}$ are usually very small. Therefore,  the only
reliable possibility of investigating the spacial relationships
between DIB carriers and simple molecules is to correlate DIB
strengths with molecular column densities derived  for an
extensive sample of sight lines. Figure~4 shows such relations for
CH  versus either  4964 or 6196 DIBs. It is clear that the 4964
DIB  is very well correlated  with the column density of CH, while
for  6196 DIB the similar correlatation is moderate (correlation
coefficients equal 0.86 and 0.48, respectively).

\begin{figure}[ht!, angle =90]
\epsfig{file=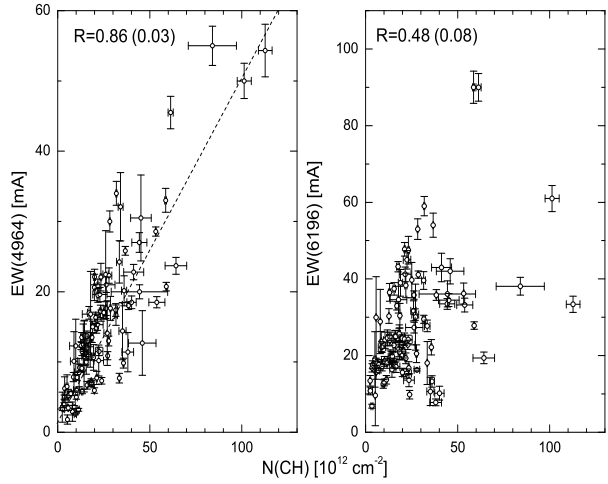, width=90mm, height=70mm, angle=0, clip=}
\caption{
Equivalent widths of 4964 and 6196 DIBs plotted vs. the column
density of CH (correlation coefficients equal 0.86 and 0.48, respectively).
Very good relation between equivalent widths of 4964 DIB and column
densities of CH molecule is presented with dotted line. }

\end{figure}

\begin{figure}[ht!, angle =90]
\epsfig{file=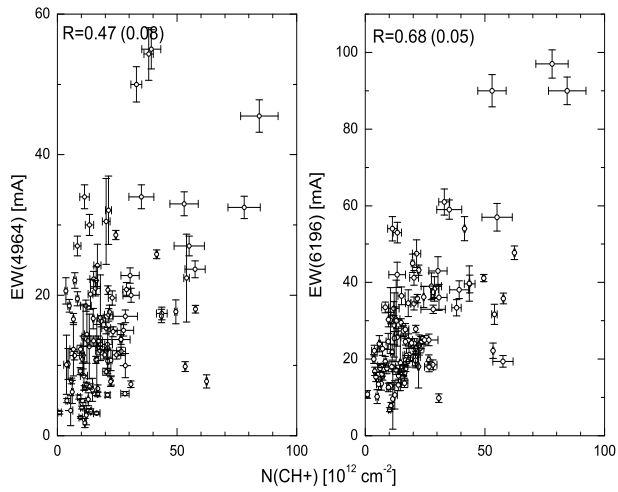,  width=90mm, height=70mm,  angle=0, clip=}
\caption{
Correlation of both 4964 and 6196 DIBs with column density of CH$^+$
(correlation coefficients equal to 0.47 and 0.68, respectively).
Note better relation
between equivalent widths of 6196 DIB and column densities of CH$^+$
molecule suggested by radial velocity shift presented in Figs. 2 and 3.  }

\end{figure}

\begin{figure}[ht!, angle =90]
\epsfig{file=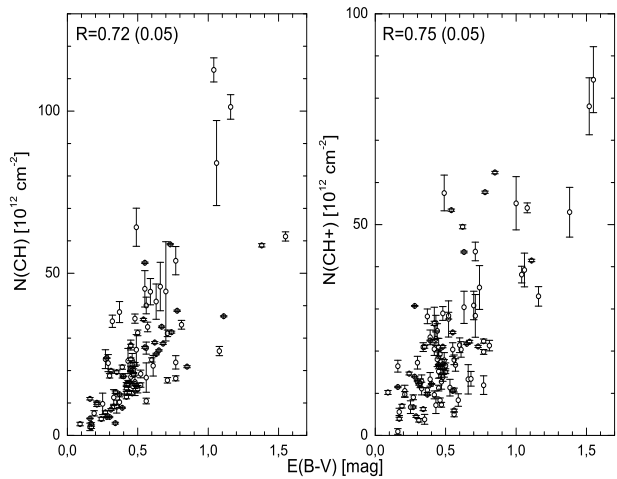, width=90mm, height=70mm, angle=0, clip=}
\caption{
Column densities of both CH and CH$^+$  correlate to a
similar degree with E(B-V) (in magnitudes). Moderate relations with correlation coefficients
equal to 0.72 and 0.75, respectively.
}
 \end{figure}

Very good relation with correlation coefficient equal to 0.86,
seen in Fig. 4 proves that the carrier of  4964 DIB occupies  the
same regions of translucent clouds as the neutral methylidyne
molecule. It also means that 74 $\%$ (0.86x0.86) of the  total
variation in equivalent width of the 4964 DIB can be explained by
the linear relationship between equivalent widths of the 4964 DIB
and column densities of CH molecule seen in Fig. 4. This molecule,
on the other hand, does not seem to share the environment with a
6196 DIB carrier. Fig. 5 shows another piece of evidence: the
column density of CH$^+$ correlates weakly with the strength of
4964 DIB (correlation coefficient equal to 0.47), and only
marginally better with that of 6196 DIB with the calculated
correlation coefficient equal to 0.68. Of note, since  CH and
CH$^{+}$ are not well mixed within the interstellar clouds, their
column density ratio cannot serve as a useful physical parameter
characterizing individual lines of sight. It seems also that a
predictive value of the colour excess E(B-V), in the discussed
contexts,  is very doubtful (Fig. 6). Both CH and CH$^+$ are
correlated with E(B-V) to a similar, moderate, degree with
correlation coefficients equal to 0.72 and 0.75, respectively.

\section{Conclusions}

The above considerations lead us to infer the following conclusions:

\begin{enumerate}
\item {
The 4964 DIB carrier is spatially correlated with neutral
methylidyne; this is confirmed by both the analysis of Doppler
velocities and by the correlation of N(CH) vs. EW(4964). The 6196
DIB, which seemed  related to CH$^{+}$  based on  Doppler shifts,
correlates only marginally better with this species than with
4964, in terms of N(CH$^{+}$) vs. EW(6196)}.

\item{ The equivalent widths of discussed diffuse bands,
as well as the molecular column densities derived here, all
correlate reasonably well with E(B-V), which makes the latter
parameter not very useful in case of finding DIB carriers originating in different
 interstellar environments.}

\end{enumerate}

\noindent
The good correlation of CH with the 4964 DIB  suggests
the existence of other, similar connections between simple
molecules and DIBs. However, in order to show the division of DIBs
between different environments, one needs the spectra of very high
S/N ratio. It could be very rewarding to check for the relations
between the strong DIBs and the interstellar molecules such as
C$_{2}$ or CO -   based on new samples, coming from homogeneous
measurements. However, a majority of DIBs  show the profiles too
broad to allow for a Doppler shift analysis. It is certainly
important to collect more spectra of the high signal-to-noise
ratio, offering the correct column density values for simple
diatomics, and thus indicating the possible relations of observed
molecules to DIBs carriers. In any case it seems to be of basic
importance to relate the origin of diffuse bands to certain
interstellar environments, defined by specific physical
parameters, the latter available from analyzes of identified
atomic and/or molecular features.

\Acknow{ JK and TW acknowledge the financial support provided by
the Polish National Center for Science for the period 2012 - 2015
(grant UMO-2011/01/BST2/05399). The authors benefited from the
funds of the Polish-Ukrainian PAS/NASU joint resarch project  No.
21(2009-2011). We are grateful for the assistance of the ESO and
BOAO observatories staff members.}

\noindent
\begin{table*}[ht!]
\tiny{

\caption{ Observational and measurement data.
Are given: HD/BD number (b -- BOAO, f -- FEROS), Spectral type / Luminosity class, E(B-V) (in magnitudes),
column densities (in 10$^{12}$ cm $^{-2}$)
of the $CH^+$ and CH molecules with error in each case,
equivalent widths (in m\AA) of the 4964 and 6196 DIBs
with error in each case.}

\vspace{-0.1cm}

\begin{center}
\begin{tabular}{rlrrrrrrrrrrr}
\hline
 HD/BD       &   Sp/L     & E(B-V)  & N($CH^+$)& $\sigma$    & N($CH$) & $\sigma$ &EW(4964) &$\sigma$&EW(6196)&$\sigma$\\
\hline
  23180  b   &  B1III      &  0.27   &  6.71    &   0.23      &   23.52 &  2.91    &  16.6   &  0.8    &  13.6  &  1.7 \\
  24398  b   &  B1Iab      &  0.29   &  4.51    &   0.23      &   22.30 &  2.61    &  10.2   &  2.2    &  15.3  &  1.2 \\
  24534  b   &  O9.5pe     &  0.56   &  4.98    &   0.58      &   40.09 &  2.57    &  18.5   &  0.9    &  10.2  &  1.8 \\
  25638  b   &  B0III      &  0.66   &  13.31   &   3.47      &   31.40 &  7.47    &  12.7   &  4.6    &  42.0  &  3.3 \\
  27778  b   &  B3V        &  0.37   &  7.58    &   0.58      &   29.46 &  3.22    &  11.4   &  2.8    &  7.8   &  0.9 \\
  34078  b   &  O9.5Ve     &  0.49   &  57.5    &   4.23      &   64.20 &  5.88    &  23.7   &  1.2    &  19.4  &  1.5 \\
  45314  f   &  O9:pe      &  0.43   &  7.18    &   1.97      &   15.44 &  0.60    &  11.4   &  0.6    &  22.7  &  1.7 \\
  46150  f   &  O6e        &  0.46   &  15.16   &   1.50      &   12.97 &  0.58    &  13.5   &  0.7    &  36.5  &  2.3 \\
  46223  f   &  O5e        &  0.48   &  28.93   &   1.62      &   21.88 &  0.60    &  20.8   &  1.0    &  38.3  &  3.2 \\
  46573  f   &  O7         &  0.63   &  30.38   &   3.80      &   41.26 &  5.44    &  22.8   &  1.1    &  43.0  &  3.7 \\
  47432  f   &  O9.5III    &  0.42   &  9.61    &   0.93      &   11.86 &  0.56    &  9.2    &  0.5    &  24.6  &  2.1 \\
  48434  f   &  B0III      &  0.21   &  10.53   &   1.39      &   10.09 &  0.52    &  5.0    &  0.3    &  13.0  &  0.9 \\
  75149  f   &  B3Ia       &  0.39   &  13.31   &   1.62      &   9.14  &  0.65    &  7.0    &  0.4    &  25.7  &  2.3 \\
  80558  f   &  B6Iae      &  0.60   &  21.41   &   1.74      &   23.28 &  0.64    &  17.2   &  0.9    &  47.5  &  3.6 \\
  93130  f   &  O6         &  0.56   &  5.90    &   0.35      &   10.60 &  0.87    &  7.8    &  0.4    &  24.0  &  2.1 \\
  94367  f   &  B9Ia       &  0.17   &  3.94    &   0.35      &   3.57  &  0.35    &  5.0    &  0.3    &  16.8  &  1.1 \\
  99264  f   &  B2IVV      &  0.27   &  9.03    &   0.58      &   7.12  &  0.43    &  5.5    &  0.3    &  17.6  &  1.4 \\
  99872  f   &  B3V        &  0.34   &  20.95   &   1.04      &   13.35 &  0.37    &  5.8    &  0.3    &  19.6  &  1.9 \\
  101191 f   &  O8         &  0.37   &  28.24   &   1.74      &   10.25 &  1.67    &  6.0    &  0.3    &  18.5  &  1.3 \\
  101205 f   &  O8v        &  0.33   &  12.85   &   1.50      &   8.92  &  0.65    &  3.6    &  0.2    &  17.2  &  1.2 \\
  101223 f   &  O7         &  0.45   &  11.23   &   0.69      &   18.03 &  1.64    &  6.8    &  0.3    &  21.8  &  2.1 \\
  104841 f   &  B2IV       &  0.09   &  10.18   &   0.46      &   3.49  &  0.54    &  4.0    &  0.2    &  6.8   &  0.5 \\
  110432 f   &  B2pe       &  0.48   &  16.09   &   0.46      &   16.51 &  2.07    &  10.8   &  0.5    &  17.1  &  1.5 \\
  116072 f   &  B2.5Vn     &  0.21   &  9.72    &   0.69      &   9.56  &  0.64    &  2.6    &  0.1    &  12.6  &  1.1 \\
  141637 f   &  B1.5V      &  0.16   &  0.93    &   0.69      &   2.61  &  1.15    &  3.3    &  0.2    &  10.8  &  1.0   \\
  147888 f   &  B3/B4V     &  0.46   &  8.33    &   0.58      &   21.84 &  0.57    &  19.5   &  1.0    &  19.3  &  1.4 \\
  147932 f   &  B5V        &  0.47   &  7.29    &   0.58      &   19.97 &  0.57    &  22.1   &  1.1    &  15.6  &  1.1 \\
  147934 f   &  B2V        &  0.45   &  15.16   &   1.62      &   27.86 &  1.53    &  22.3   &  1.1    &  16.3  &  0.3 \\
  148379 f   &  B2Iab      &  0.77   &  19.79   &   0.58      &   22.53 &  2.06    &  11.5   &  0.6    &  45.0  &  2.4 \\
  148688 f   &  B1Ia       &  0.52   &  27.63   &   4.28      &   18.98 &  1.18    &  15.0   &  0.8    &  39.0  &  3.5 \\
  148937  b  &  O6e        &  0.61   &  20.39   &   1.85      &   21.47 &  3.18    &  15.3   &  0.8    &  41.2  &  1.9 \\
  149404  b  &  O9Ia       &  0.71   &  43.63   &   2.20      &   31.50 &  1.02    &  17.4   &  0.9    &  39.6  &  2.3 \\
  150136  b  &  O5         &  0.42   &  20.49   &   1.50      &   13.14 &  0.36    &  9.1    &  0.5    &  24.0  &  1.3 \\
  150168  b  &  B1Iab/Ib   &  0.16   &  16.43   &   1.39      &   11.31 &  0.38    &  3.2    &  0.2    &  13.7  &  1.1 \\
  152076  b  &  B0.5III    &  0.43   &  22.92   &   1.62      &   22.94 &  3.86    &  19.6   &  1.0    &  22.8  &  1.2 \\
  152218  b  &  O9V        &  0.45   &  18.29   &   1.85      &   20.70 &  0.66    &  16.3   &  0.8    &  21.1  &  1.7 \\
  152219  b  &  O9IV       &  0.44   &  24.77   &   1.74      &   13.27 &  0.60    &  11.5   &  0.6    &  25.0  &  1.1 \\
  152234  b  &  B0.5Ia     &  0.42   &  26.58   &   3.80      &   15.88 &  0.43    &  13.7   &  0.7    &  25.0  &  1.5 \\
  152246  b  &  O9Ib       &  0.44   &  18.06   &   1.74      &   16.31 &  0.31    &  13.0   &  0.7    &  25.0  &  1.6 \\
  152314  b  &  B0.5Iab    &  0.39   &  23.26   &   1.74      &   21.05 &  0.66    &  14.8   &  0.7    &  24.3  &  1.6 \\
  152667  b  &  B0.5Iae    &  0.47   &  12.73   &   1.74      &   18.64 &  0.50    &  13.5   &  0.7    &  30.4  &  3.2 \\
  154090  b  &  B1Iae      &  0.45   &  20.95   &   2.08      &   27.63 &  0.45    &  13.0   &  0.7    &  20.5  &  2.3 \\
  164794  b  &  O4V        &  0.33   &  11.11   &   1.50      &   11.67 &  2.47    &  8.5    &  2.0    &  18.0  &  1.3 \\
  165688  b  &  WN         &  1      &  55.06   &   6.33      &   --    &  --      &  27.0   &  1.4    &  57.0  &  3.6 \\
  166734  b  &  O8e        &  1.38   &  52.95   &   5.91      &   58.60 &  0.52    &  33.0   &  1.7    &  90.0  &  4.2 \\
  168075  b  &  O7V        &  0.74   &  35.07   &   5.21      &   31.85 &  0.30    &  34.0   &  1.7    &  59.0  &  2.5 \\
  168137  b  &  O8.5V      &  0.68   &  13.43   &   1.74      &   28.29 &  0.28    &  30.0   &  1.5    &  53.0  &  2.7 \\
  168607  b  &  B9Ia       &  1.55   &  84.38   &   7.80      &   61.36 &  1.41    &  45.5   &  2.3    &  90.0  &  3.6 \\
  168625  b  &  B6Ia       &  1.52   &  78.05   &   6.75      &   --    &  --      &  32.5   &  1.6    &  97.0  &  3.7 \\
  184915  b  &  B0.5III    &  0.19   &  6.94    &   0.46      &   6.85  &  0.81    &  ---    &  ---    &  16.1  &  0.6 \\
  190603  b  &  B1.5Iae    &  0.71   &  28.36   &   5.06      &   17.00 &  0.85    &  17.0   &  0.9    &  33.0  &  1.1 \\
  203064  b  &  O8V        &  0.25   &  6.71    &   1.50      &   9.75  &  3.30    &  12.3   &  3.8    &  18.3  &  0.5 \\
  203938  b  &  B0.5IV     &  0.7    &  30.80   &   3.38      &   34.20 & 15.30    &  20.0   &  1.0    &  36.1  &  3.4 \\
  204827  b  &  B0V        &  1.06   &  39.24   &   4.01      &   84.00 & 13.10    &  55.0   &  2.8    &  38.1  &  2.3 \\
  206165  b  &  B2Ib       &  0.46   &  19.33   &   1.16      &   23.09 &  2.21    &  ---    &  ---    &  24.2  &  1.1 \\
  206267  b  &  O6         &  0.49   &  11.33   &   2.08      &   26.40 &  4.71    &  21.0   &  1.1    &  27.3  &  1.4 \\
  207198  b  &  O9II       &  0.55   &  20.37   &   1.62      &   35.64 &  5.62    &  30.5   &  6.1    &  34.5  &  1.7 \\
  207538  b  &  B0V        &  0.59   &  8.39    &   1.50      &   44.34 &  4.07    &  27.0   &  1.4    &  33.5  &  1.4 \\
  208501  b  &  B8Ib       &  0.77   &  11.92   &   2.20      &   53.86 &  4.36    &  18.5   &  0.8    &  33.2  &  1.8 \\
  209481  b  &  O9V        &  0.35   &  3.82    &   1.16      &   8.87  &  2.01    &  10.1   &  4.2    &  22.2  &  1.4 \\
  210839  b  &  O6e        &  0.56   &  10.76   &   0.46      &   26.93 &  1.86    &  14.1   &  4.6    &  30.7  &  1.5 \\
  213087  b  &  B0.5Ibe    &  0.56   &  17.94   &   1.16      &   17.84 &  4.50    &  16.8   &  6.1    &  34.6  &  3.5 \\
  326331  b  &  B          &  0.3    &  17.25   &   1.50      &   18.38 &  0.68    &  13.5   &  0.7    &  20.9  &  1.9 \\
  BD134927 f &  O7II       &  1.16   &  32.99   &   2.31      &   101.3 &  3.80    &  50.0   &  2.5    &  61.0  &  3.4 \\
  BD134930 f &  O9.5V      &  0.53   &  11.34   &   1.97      &   --    &  --      &  34.0   &  1.7    &  54.0  &  3.2 \\
  CPD417733 b &  O9III     &  0.49   &  17.82   &   1.85      &   14.42 &  0.65    &  12.1   &  0.6    &  19.4  &  1.5 \\
  CPD417735 b &  O9V       &  0.43   &  26.62   &   1.85      &   15.97 &  0.97    &  12.0   &  0.6    &  18.0  &  1.3 \\
  \hline
\end{tabular}
\end{center}
}
\end{table*}

\newpage
\begin{table*}[ht!]
\tiny{

\caption{ Observational and measurement data (continued).
Are given: HD/BD number (H -- HARPS, U -- UVES), Spectral type / Luminosity class, E(B-V) (in magnitudes),
column densities (in 10$^{12}$ cm $^{-2}$)
of the $CH^+$ and CH molecules with error in each case, equivalent widths
(in m\AA) of the 4964 and 6196 DIBs
with error in each case.}

\vspace{-0.1cm}

\begin{center}
\begin{tabular}{rlrrrrrrrrrrr}
\hline
 HD/BD       &   Sp/L     & E(B-V)  & N($CH^+$)& $\sigma$    & N($CH$) & $\sigma$ &EW(4964) &$\sigma$&EW(6196)&$\sigma$\\
\hline
   144217 H   &  B0.5V      &  0.17   &  5.582   &   0.85      &   2.481 & 0.35     &  3.58   &  2.1    &  13.4  &  1.4  \\
  147165 H   &  B1III      &  0.34   &  6.249   &   0.36      &   3.808 & 0.27     &  6.24   &  1.6    &  17.6  &  1.0    \\
  147889 H   &  B2III/IV   &  1.04   &  38.14   &   1.96      &  112.70 & 3.71     &  54.3   &  3.7    &  33.4  &  2.1   \\
  148184 H   &  B2Vnec     &  0.48   &  13.82   &   0.97      &   36.01 & 1.41     &  20.1   &  2.1    &  13.3  &  1.5  \\
  152233 H   &  O6III      &  0.42   &  26.58   &   0.48      &   14.55 & 0.51     &  13.7   &  2.2    &  19.8  &  1.3   \\
  152235 H   &  B0.7Ia     &  0.78   &  57.71   &   0.27      &   38.42 & 0.20     &  18.0   &  0.6    &  35.8  &  1.4  \\
  152249 H   &  O9Ib       &  0.48   &  21.04   &   0.27      &   13.84 & 1.57     &  12.5   &  3.0    &  19.7  &  1.4   \\
  163800 H   &  O7         &  0.57   &  16.73   &   1.48      &   33.46 & 1.48     &  24.2   &  3.0    &  27.7  &  1.5   \\
  179406 H   &  B3V        &  0.31   &  3.598   &   0.43      &   19.95 & 0.44     &  20.6   &  1.9    &  20.0  &  1.1   \\
  147933 H   &  B2V        &  0.45   &  16.25   &   0.47      &   23.46 & 0.44     &  22.1   &  2.0    &  16.7  &  1.7   \\
  110432 U   &  B2pe       &  0.48   &  16.88   &   0.10      &   15.92 & 1.00     &  6.7    &  0.6    &  17.5  &  0.7  \\
  152236 U   &  B1Iape     &  0.65   &  21.75   &   0.20      &   26.22 & 0.21     &  17.6   &  1.0    &  35.8  &  1.1  \\
  152424 U   &  O9Iab      &  0.63   &  43.49   &   0.21      &   25.09 & 0.25     &  17.0   &  1.0    &  39.7  &  4.6   \\
   35149 U   &  B2V        &  0.16   &  11.48   &   0.14      &   5.23  & 0.12     &  1.8    &  0.7  &  9.6   &  7.8   \\
   37903 U   &  B2V        &  0.31   &  12.12   &   0.19      &   7.97  & 0.25     &  2.9    &  1.4  &  28.9  &  8.4    \\
   52266 U   &  O9V        &  0.24   &  14.67   &   0.33      &   5.07  & 0.54     &  6.5    &  1.6  &  19.0  &  7.2    \\
   52382 U   &  B1Ib       &  0.39   &  22.51   &   0.12      &   8.53  & 0.16     &  7.8    &  0.6  &  23.7  &  2.6    \\
   53974 U   &  B0.5IV     &  0.28   &  13.93   &   0.10      &   5.67  & 0.12     &  3.5    &  0.4  &  15.7  &  1.8    \\
   58510 U   &  B1Ia       &  0.30   &  12.89   &   0.15      &   5.67  & 0.19     &  5.2    &  0.9  &  30.0  &  10.6    \\
   73882 U   &  O8V        &  0.67   &  22.25   &   0.32      &   33.48 & 0.22     &  7.7    &  0.7  &  18.1  &  5.6    \\
   75149 U   &  B3Ia       &  0.40   &  12.21   &   0.12      &   9.14  & 0.11     &  7.2    &  0.4  &  25.4  &  1.7    \\
   76341 U   &  O9III      &  0.54   &  53.43   &   0.28      &   35.74 & 0.35     &  9.9    &  0.7  &  22.2  &  2.0    \\
   91452 U   &  B0III      &  0.55   &  10.54   &   0.21      &   27.16 & 0.26     &  10.9   &  0.6  &  30.1  &  8.6    \\
   92964 U   &  B2.5Iae    &  0.37   &  9.647   &   0.12      &   12.69 & 0.16     &  12.3   &  0.6  &  30.3  &  1.9    \\
  148379 U   &  B2Iab      &  0.77   &  22.37   &   0.48      &   17.60 & 0.77     &  10.7   &  1.5  &  43.2  &  1.4    \\
  149404 U   &  O9Ia       &  0.62   &  49.50   &   0.46      &   28.63 & 0.32     &  17.6   &  1.7  &  41.1  &  0.9   \\
  151932 U   &  WN7        &  0.50   &  15.01   &   0.34      &   31.61 & 0.87     &  16.7   &  1.4  &  29.5  &  1.0    \\
  154445 U   &  B1V        &  0.35   &  21.02   &   0.31      &   19.53 & 0.40     &  10.6   &  1.8  &  22.8  &  0.8  \\
  148688 U   &  B1Ia       &  0.52   &  28.39   &   0.80      &   15.54 & 0.69     &  10.0   &  1.7  &  37.4  &  1.6    \\
  152270 U   &  WC7        &  0.50   &  20.68   &   0.69      &   14.27 & 1.15     &  10.8   &  5.1  &  16.4  &  2.0    \\
  115363 U   &  B1Ia       &  0.81   &  21.32   &   1.24      &   34.14 & 1.36     &  32.1   &  4.9  &  ---   &   ---    \\
  136239 U   &  B2Iae      &  1.08   &  53.97   &   1.15      &   25.63 & 1.42     &  20.5   &  6.3  &  31.7  &  2.6    \\
  154368 U   &  O9.5Iab    &  0.73   &  21.11   &   0.26      &   58.94 & 0.15     &  20.7   &  0.6  &  27.8  &  0.9  \\
  157038 U   &  B4Ia       &  0.85   &  62.35   &   0.29      &   21.20 & 0.31     &  7.7    &  0.9  &  47.7  &  1.8   \\
  161056 U   &  B3Vn       &  0.55   &  24.43   &   0.26      &   53.26 & 0.20     &  28.6   &  0.6  &  36.2  &  2.7   \\
  169454 U   &  B1Ia       &  1.11   &  20.91   &   0.28      &   36.74 & 0.17     &  25.8   &  0.6  &  54.0  &  3.2   \\
  170740 U   &  B2V        &  0.45   &  16.80   &   0.11      &   19.86 & 0.10     &  6.0    &  0.4  &  24.5  &  0.9  \\
  149757 U   &  O9.5V      &  0.28   &  30.71   &   0.12      &   23.86 & 0.07     &  7.33   &  0.5  &  9.8   &  1.1   \\
  210121 U   &  B3V        &  0.32   &  12.36   &   1.04      &   35.20 & 1.91     &  14.4   &  4.1  &  10.7  &  3.7    \\
\hline
\end{tabular}
\end{center}
}
\end{table*}

\end{document}